\documentclass[aps,prl,twocolumn,notitlepage,superscriptaddress,showpacs,nofootinbib]{revtex4-2}

\usepackage{amstext,graphicx,times}
\usepackage[dvipsnames]{xcolor}
\usepackage{enumerate}
\usepackage{amsmath}
\usepackage{amsthm}
\usepackage{amssymb}
\usepackage{dsfont}
\usepackage{amsmath}
\usepackage{amssymb}
\usepackage{amsfonts}
\usepackage{enumerate}
\usepackage{mathrsfs}
\usepackage{graphicx}
\usepackage{epstopdf}
\usepackage{enumerate}

\newcommand{\tr}{\text{Tr}}

\usepackage[
colorlinks,
linkcolor = blue,
citecolor = blue,
urlcolor = blue]{hyperref}

\newcommand{\kommentar}[1]{}

\makeatletter
\usepackage{amsmath, amsfonts, amssymb, amsthm, mathrsfs, amssymb, braket, bbm, bm,graphicx,times,txfonts,color,subfigure}
\usepackage{hyperref}

\theoremstyle{plain}

\theoremstyle{plain}

\theoremstyle{definition}

\theoremstyle{plain}

\makeatother

\providecommand{\examplename}{Example}
\providecommand{\propositionname}{Proposition}
\providecommand{\theoremname}{Theorem}
\providecommand{\lemmaname}{Lemma}

\begin{document}
	\title{ Simultaneous Detection of High-Dimensional Entanglement for Two Unknown Quantum States }

  \author{Mao-Sheng Li}
 
 \affiliation{ School of Mathematics,
 	South China University of Technology, Guangzhou
 	510641,  China}

 \author{Chang-Yue Zhang}
 \affiliation{ College of Information Science and Technology,    Jinan University, Guangzhou  510632, China}

 \author{Zheng Zheng}\affiliation{School of Mathematics, Guangdong University of Education, Guangzhou 510303,   China}
 
 \author{Zhihua Chen}
 
 \affiliation{School of Science, Jimei University, Xiamen 361021,China}

 \author{Zhen-Peng Xu}
 \email{zhen-peng.xu@ahu.edu.cn}
 \affiliation{School of Physics and Optoelectronics Engineering,
 	Anhui University, Hefei 230601, China}

 \author{Zhihao Ma}
 \email{mazhihao@sjtu.edu.cn}
 \affiliation{School of Mathematical Sciences, MOE-LSC, Shanghai Jiao Tong University, Shanghai 200240,
 	China}
 \affiliation{Shanghai Seres Information Technology Co., Ltd, Shanghai 200040, China}
 \affiliation{Shenzhen Institute for Quantum Science and Engineering, Southern University of Science and Technology, Shenzhen 518055, China}
 
 \author{Yan-Ling Wang}
 
 \email{ylwang@dgut.edu.cn}
 \affiliation{ School of Computer Science and Technology, Dongguan University of Technology, Dongguan, 523808, China}

 \author{Shao-Ming Fei} 
 \email{smfei@mis.mpg.de}
 \affiliation{
 	School of Mathematical Sciences, Capital Normal University, Beijing 100048, China}
 \affiliation{Max-Planck-Institute for Mathematics in the Sciences, 04103 Leipzig, Germany
 }
 
 \author{Zhu-Jun Zheng}\email{zhengzj@scut.edu.cn}
 \affiliation{ School of Mathematics,
 	South China University of Technology, Guangzhou
 	510641,  China}
 
 \author{Otfried G\"{u}hne}
 \email{otfried.guehne@uni-siegen.de}
 \affiliation{Naturwissenschaftlich-Technische Fakultät, Universität Siegen, Walter-Flex-Straße 3, 57068 Siegen, Germany}
\begin{abstract}
The state overlap, quantified via 
	$\tr[\rho \sigma]$, is a metric widely used to assess the closeness between two quantum states $\rho$ and $\sigma$.
  Although global state overlap alone does not directly capture entanglement properties, we uncover  that incorporating local state overlaps provide  profound insights into the entanglement characteristics of quantum states. To be precise, the ratio of global to local state overlaps provides a lower bound on the Schmidt number, which is usually used for quantifying high-dimensional entanglement. Unlike conventional methods for detecting entanglement, the approach here can simultaneously reveal entanglement information for two unknown quantum states. Moreover, state overlap can be efficiently determined through local randomized measurement methods, which ensures the experimental feasibility of our approach.   In a special case, our criterion  reduces to an  entanglement criterion that is more powerful than the two  criteria used most in experiment—the purity criterion and the fidelity-based criterion and   also outperform the $p_3$-PPT method in specific instances.  Our findings highlight a promising direction for advancements in   entanglement detection experiments.

\end{abstract}

\maketitle

\emph{Introduction.---} Entanglement, a fascinating cornerstone of quantum mechanics, is pivotal in the realm of quantum information processing  \cite{AmicoEntangle08,HHHH09,Guhne09}.  It serves as the indispensable asset underpinning   quantum teleportation \cite{Teleportation}, quantum key distribution \cite{QKD,QKD_Curty04,QKD_Pan20}, quantum metrology  \cite{QM_Giovannetti04,QM_Giovannetti11,QM_Degen17} and  the verification of quantum nonlocality \cite{BellNL_Brunner14}.  Consequently, the detection and quantification of entanglement are crucial steps for driving forward  the progress of quantum technologies.

The study of entanglement traces its roots back to the groundbreaking work of Einstein, Podolsky, and Rosen (EPR) \cite{EPR}, as well as Schrödinger’s contributions \cite{Schrodinger}.  Decades later, Peres introduced the widely recognized positive partial transpose (PPT) criterion \cite{CC_PPT_Peres96}, followed by a proliferation of entanglement criteria, including the reduction criterion, majorization criterion, cross-norm or realignment criterion, covariance matrices criterion, local uncertainty criterion, etc. \cite{CC_Quantify_Vedral97,CC_Reduction_Cerf99,CC_Reduction_HH_99,CC_TM_Major_Nielsen01,CC_TM_Majorization_Hiroshima03,CC_RM_Chen03,CC_GRC_Albeverio03,CC_CCNC_Rudolph03,CC_LUR_Guhne06,CC_CM_Guhne07,CC_EW_Chruscinski14,TM_Trace_Huber24}. A conventional approach to detect entanglement involves state tomography to reconstruct the density matrix, followed by the application of an appropriate criterion. However, the requirement of measurement for tomography grow exponentially with the number of qubits, rendering it impractical for many experimental settings. Consequently, the development of entanglement detection methods that are amenable to physical realization has attracted a major research endeavor \cite{EX_IPE_Flammia11,EX_ME_Guhne04,EX_EM_Guhne07,EX_Theory_Friis19,NewD_HDE_Tavakoli23}.  A popular method for certifying entanglement is to measure the fidelity  $\langle \Psi|\rho|\Psi\rangle$ between an unknown state $\rho$ and a maximally entangled state $|\Psi\rangle$ \cite{EX_FBC_Fickler14}. A sufficiently high fidelity indicates entanglement, and this approach is known as the fidelity-based criterion \cite{NewD_FBC_Guhne21}.   Yet, there are entangled states that elude detection by this criterion, termed unfaithful states \cite{FBC_Riccardi21,NewD_FBC_Guhne21,NewD_FBC_Weilenmann20}. 
Beyond merely confirming the existence of entanglement, determining its dimensionality is crucial.  High-dimensional entangled states, for instance, have been utilized to enhance security bounds in quantum communication \cite{HD_Cozzolino19}. The significance of this is highlighted by numerous theoretical and experimental findings related to high-dimensional entanglement (see Refs. \cite{Schmidt_Terhal_00,EX_HD_Laskowski12,EX_HDE_Photonic19,NewD_HDE_Schmidt_Huber18,NewD_HDE_Schmidt_Bavaresco18,EX_HD_Hu21,NewD_PTM_Yu21,NewD_HDE_Tavakoli23,NewD_HDE_He2023,EX_HD_Euler23,NewD_SM_Zhang24}). The $r$-fidelity-based criterion ($r$-FBC) derived from \cite{EX_FBC_Fickler14}, holds particular significance in this field.    Nevertheless, it remains intriguing to explore  whether there are other methods capable of detecting entangled states that $r$-FBC may fail to identify.

The emergence of randomized measurement techniques has made it possible to empirically study entanglement by examining the contrast between global and local purities, such as $\tr[\rho^2]$, or through the lens of second order Rényi entropies, $-\log_2 \tr[\rho^2]$ \cite{CC_Pure_HH96,EX_EE_Islam15,EX_purity_Ekert02}. Furthermore, recent insights suggest that entanglement can be detected by scrutinizing the moments of the partial transpose, with the $p_3$-PPT criterion serving as a straightforward illustration, facilitated by randomized measurements \cite{EX_RM__Renyi_Brydges19,NewD_RM_Elben19,NewD_RM_Knips20,NewD_RM_P3PPT_Elben20,Random24, NewD_PTM_Yu21}. Nonetheless, the hunt for additional measurable quantities through these innovative randomized measurement methods—quantities that could serve as indicators of entanglement or even reveal high-dimensional entanglement—remains an intriguing challenge.

  The state overlap,  defined as $\tr[\rho \sigma]$ (also known as the Hilbert-Schmidt inner product of $\rho$ and $\sigma$), offers insight into the similarity   between two quantum states. Recently, research has found that this quantity can be measured through local randomized measurements, leading to the proposal of a protocol for cross-platform verification of quantum simulators and quantum computers \cite{IPC_Elben20}.  A natural question arises: Can such a fundamental quantity also offer insights into the entanglement properties of quantum states?  In this work, we answer this in the affirmative by demonstrating that the ratio of global to local state overlap not only provides information about entanglement but can also be utilized to detect the Schmidt number, which quantifies high-dimensional entanglement.  Additionally, our method facilitates the simultaneous determination of the Schmidt numbers for two distinct quantum states, highlighting its uniqueness compared to known methods.

\vskip 4pt

\emph{Entanglement and Schmidt number.---}Let $\mathcal{H}_{A}$ and $\mathcal{H}_{B}$ be Hilbert spaces of dimension $d_A$ and $d_B$  with computational basis  $\{|0\rangle_A,|1\rangle_A,\cdots, |d_A-1\rangle_A\}$ and    $\{|0\rangle_B,|1\rangle_B,\cdots, |d_B-1\rangle_B\}$ respectively.  Let $\mathcal{H}=\mathcal{H}_A \otimes \mathcal{H}_B$ be the composed system of $A$ and $B$.   
Denote $\mathbb{L} (\mathcal{H} )$ the set of all linear operators acting on $\mathcal{H}$ and     $\mathbb{D} (\mathcal{H} )$ the set of all  the density matrices of the system $\mathcal{H}$, i.e.,  the set of all selfadjoint positive semidefinite operators with trace 1 from $\mathcal{H}$ to itself. We  denote the partial trace operator   $\tr_{A}$   to be the linear operation from  $\mathbb{L}(\mathcal{H})  $ to $\mathbb{L} (\mathcal{H}_{B} )$   by sending $P\otimes Q$ to $\tr[P] Q$.  Similarly, we   denote $\tr_{B}$   a linear operator from  $\mathbb{L}(\mathcal{H})  $ to $\mathbb{L} (\mathcal{H}_{A} )$  by sending $P\otimes Q$ to $\tr[Q]P$.

A pure state $|\Phi\rangle$ in $\mathcal{H}$ is just a unit vector.  It is well known that each bipartite pure state can be written as the form $|\Phi\rangle=\sum_{k=1}^r \sqrt{\lambda_k} |e_k f_k\rangle$ where $\lambda_1\geq \cdots \geq \lambda_r>0$, $\{|e_k\rangle\}_k$ and $\{|f_k\rangle\}_k$ are orthonormal sets of subsystems $A$ and $B$ respectively. The number  $r$ is known as the Schmidt number of $|\Phi\rangle$ and denoted as $\mathrm{SN}(|\Phi\rangle)=r.$ For a bipartite mixed state $\rho\in \mathbb{D}(\mathcal{H})$, the Schmidt number of $\rho$ is defined as (see also in Ref. \cite{Schmidt_Terhal_00}), 
\begin{equation}\mathrm{SN}(\rho)=\min_{ \mathcal{D}(\rho)} \max_{|\phi_i\rangle\in \mathcal{D}(\rho) } \mathrm{SN}(|\phi_i\rangle),
\end{equation} 
where  $\mathcal{D}(\rho):=\{\{|\phi_i\rangle\}_{i\in I} \ \big|\   \rho= \sum_{i\in I} p_i |\phi_i\rangle\langle \phi_i| \}$ denotes all the pure state decompositions of $\rho$.  The Schmidt number gives a characterization for the dimensionality of entanglement. $\rho$ is entangled if $\mathrm{SN}(\rho)\geq 2$, and separable otherwise.
	
	\vskip 8pt

	\emph{Inner product criterion for Schmidt number.---} The purity criterion (PC) is   a useful tool in experimental settings to detect entanglement.   The criterion states that for a bipartite state $\rho\in \mathbb{D}(\mathcal{H}_A\otimes \mathcal{H}_B)$, if   $\tr[\rho^2]>\min\{\tr[\rho_A^2], \tr[\rho_B^2]\}
$, then $\rho$ must be entangled \cite{CC_Pure_HH96,EX_EE_Islam15}. The Hilbert-Schmidt inner product of two matrices  $M,N$ of the same dimension  is defined as $\langle M,N\rangle:=\tr[M^\dagger  N ].$ Under this notation, the above inequality can be written as 
	$\langle \rho, \rho\rangle >\min\{\langle \rho_A, \rho_A\rangle, \langle \rho_B, \rho_B\rangle\}.$ Inspired by the PC,   
  we obtain our main result to  determine the high dimensional  entanglement, i.e., the Schmidt number, which works for two states simultaneously.

	\emph{Theorem 1.---}(Inner product criterion for Schmidt number) 	Let $\rho\in \mathbb{D}(\mathcal{H}_A\otimes\mathcal{H}_B)$ be   a bipartite density matrix. If  $\mathrm{SN}(\rho) \leq r$, then for all  $\sigma\in \mathbb{D}(\mathcal{H}_A\otimes\mathcal{H}_B)$, we have  
	\begin{equation}\label{eq:Inner_ineq_SCHM}
		\langle \rho,\sigma\rangle\leq \min\{ r \ \langle \rho_A,\sigma_A\rangle,r\ \langle \rho_B,\sigma_B\rangle\}
	\end{equation}
	As a consequence, if the inequality \eqref{eq:Inner_ineq_SCHM} is violated, then both $\rho$ and $\sigma $ are of Schmidt numbers at least $(r+1)$.

  The proof of Theorem 1 is given in Appendix A. For a fixed $r$, we refer to the corresponding criterion as the $r$-inner product criterion ($r$-IPC).
	
	A widely used approach for Schmidt number detection is the well-known $r$-reduction criterion ($r$-RC), which states that if $\mathrm{SN}(\rho) \le r$, then
	\begin{equation}\label{eq:r_RC}
		r  \mathbb{I}_A \otimes \rho_B - \rho \ge \mathbf{0},
		\qquad
		r  \rho_A \otimes \mathbb{I}_B - \rho \ge \mathbf{0}.
	\end{equation}	
	The following observation establishes a direct connection between these two criteria.

	\emph{Observation 1.---}(Equivalence of $r$-IPC and $r$-RC) A bipartite state $\rho\in \mathbb{D}(\mathcal{H}_A\otimes\mathcal{H}_B)$ violates the inequality \eqref{eq:Inner_ineq_SCHM} for some state $\sigma$ if and only if it violates the inequalities \eqref{eq:r_RC}.

	The proof of Observation 1 is given in Appendix B. Observation 1 shows that the $r$-IPC and the $r$-RC possess exactly the same detectability for the Schmidt number, in the sense that they certify entanglement for the same set of states.	It is important to emphasize that this equivalence concerns only the detection power, and does not imply that the two criteria share the same operational or experimental properties. In fact, to the best of our knowledge, no physically feasible experimental implementation of the $r$-RC is currently available.
	
It is important to stress, however, that this equivalence concerns only
the detection power and does not imply identical operational or experimental properties.
In particular, to the best of our knowledge, no physically feasible experimental
implementation of the $r$-RC is currently available.
By contrast, as we demonstrate below, the $r$-IPC admits a natural operational
realization and offers additional information-processing advantages.
These advantages become manifest once Theorem~1 is reformulated
in a quantitative and experimentally accessible form. 

To this end, we introduce a ratio that directly converts
\eqref{eq:Inner_ineq_SCHM} into lower bounds on Schmidt numbers. For each $X \in {A,B}$, we define the ratio of global to local state overlap as
$	\mathcal{S}_X(\rho, \sigma) := \displaystyle\frac{\langle \rho, \sigma \rangle}{\langle \rho_X, \sigma_X \rangle}
$
if \(\langle \rho_X, \sigma_X \rangle \neq 0\), and  set \(\mathcal{S}_X(\rho, \sigma) = 0\) otherwise.
Let
$
\mathcal{S}(\rho,\sigma)
=
\max\!\left\{
\mathcal{S}_A(\rho,\sigma),
\mathcal{S}_B(\rho,\sigma)
\right\}.
$
Then Theorem~1 immediately implies
\begin{equation}\label{eq:Schmidt_estimation}
	\mathrm{SN}(\rho)\ge \lceil \mathcal{S}(\rho,\sigma)\rceil,
	\qquad
	\mathrm{SN}(\sigma)\ge \lceil \mathcal{S}(\rho,\sigma)\rceil.
\end{equation}

This reformulation reveals that the $r$-IPC not only matches the
detectability of the $r$-RC, but also provides a quantitative and operationally
meaningful estimator for the Schmidt number.
In the following, we highlight several advantages of the $r$-IPC,
summarized in points (i)--(vi).

(i) According to Theorem 1, the $r$-IPC can simultaneously detect the Schmidt numbers of two quantum states  $\rho$ and $\sigma$. Let us illustrate this with the following example.

\begin{figure}[t!]
	\centering
	\includegraphics[scale=0.7]{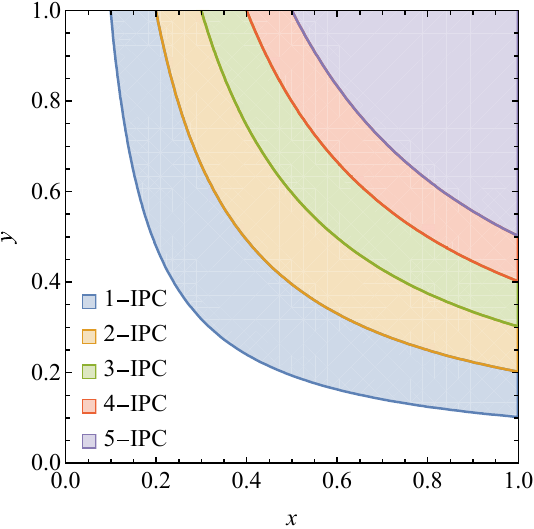}	
	\caption{This figures shows the points $(x,y)$ where the Schmidt number of $\rho_{\mathrm{Iso}}(x)$ and $\rho_{\mathrm{Iso}}(y)$   can be detected via $r$-IPC ($r=1,2,3,4,5$ for $d=10$). That is, the corresponding $\rho_{\mathrm{Iso}}(x)$ and $\rho_{\mathrm{Iso}}(y)$ violate the inequality \eqref{eq:Inner_ineq_SCHM}.  }\label{fig:compare_Isoxy}
\end{figure}

\emph{Example 1.---}  We consider the isotropic states which are a class of $U\otimes U^*$ invariant mixed states in $\mathcal{H}_{A}\otimes \mathcal{H}_{B}$ with $d_A=d_B=d$, that is,
\begin{equation} \rho_{\mathrm{Iso}}(x)=\frac{1-x}{d^2-1} \mathbb{I}_{d^2} +\frac{d^2x-1}{d^2-1}|\Psi\rangle\langle \Psi|
\end{equation}
where $|\Psi\rangle$ is the maximally entangled states $\frac{1}{\sqrt{d}} \sum_{i=0}^{d-1} |ii\rangle$. Let us consider two isotropic states, $\rho_{\mathrm{Iso}}(x)$ and $\rho_{\mathrm{Iso}}(y)$. To certify the conditions that they violate the inequality \eqref{eq:Inner_ineq_SCHM}, we plot the area of the points $(x,y)$ in which the Schmidt numbers of $\rho_{\mathrm{Iso}}(x)$ and $\rho_{\mathrm{Iso}}(y)$ can be detected via $r$-IPC (see Fig. \ref{fig:compare_Isoxy}). 
Particularly, setting $y=1$, one finds that $\mathcal{S}(\rho_{\mathrm{Iso}}(x),\rho_{\mathrm{Iso}}(1))=dx$. Therefore, 
$\mathrm{SN}(\rho_{\mathrm{Iso}}(x))\geq \lceil  dx \rceil$ which coincide with  the result in Ref. \cite{Schmidt_Terhal_00}.
 Moreover, Fig. \ref{fig:compare_Isoxy} can be used to demonstrate the robustness of our method against variations in the range of parameters.

\begin{figure} 
	\centering
	\includegraphics[scale=0.34]{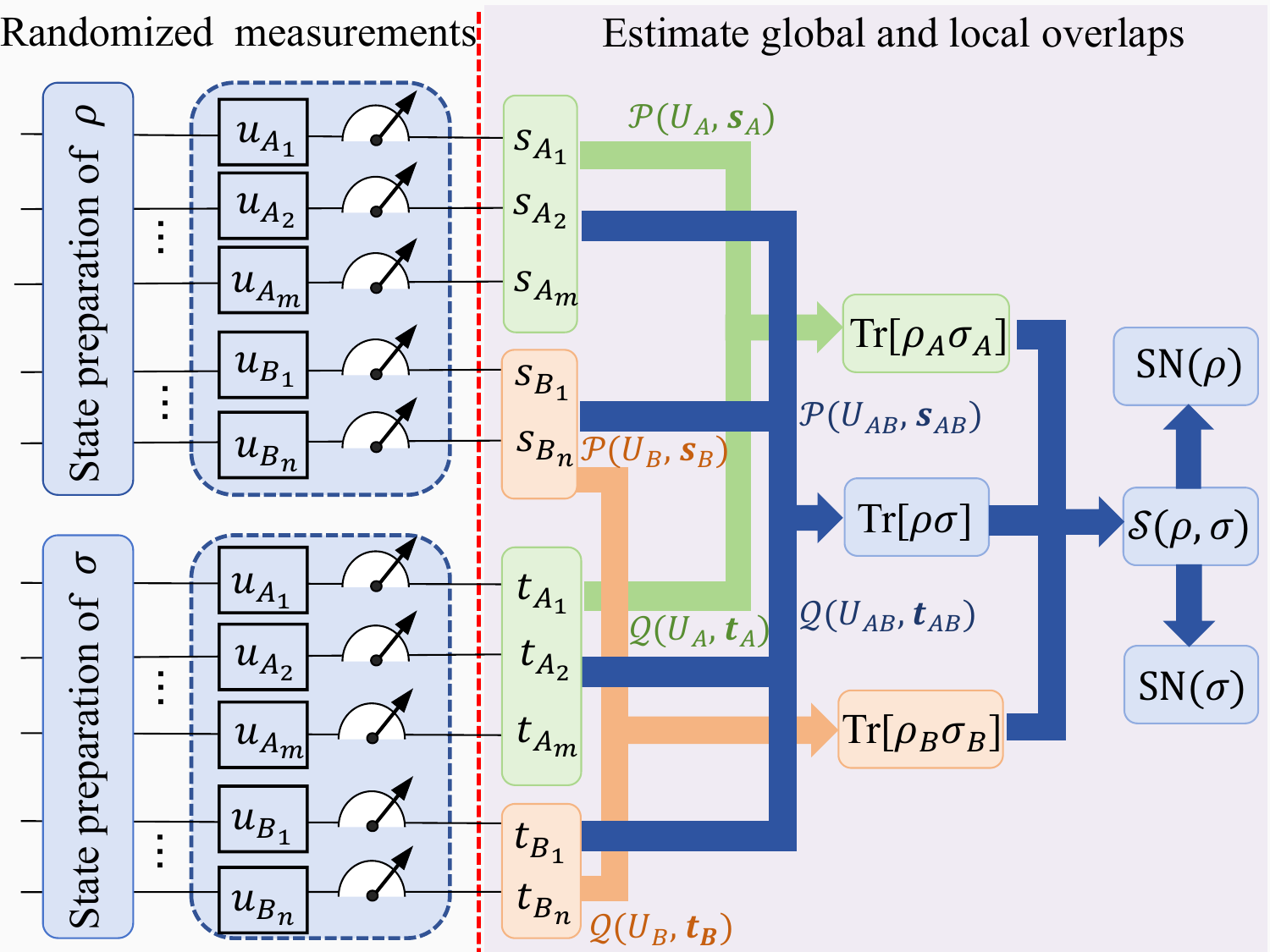}	
	\caption{This figure shows how to use local randomized measurements  to detect   the Schmidt number. In view of our current context, we propose a slight modification to the existing methods for estimating state overlap through local randomized measurements (the full details of the method can be found in Ref. \cite{IPC_Elben20}). Consider the simplest case where   $\mathcal{H}_A=\otimes_{i=1}^m \mathcal{H}_{A_i}$ and  $\mathcal{H}_B=\otimes_{j=1}^n \mathcal{H}_{B_j}$ with each   $\mathcal{H}_{A_i}$ and $\mathcal{H}_{B_j}$ having a dimension of $\ell \geq 2$. We  apply the same unitary $U_{AB}=U_A\otimes U_B$, where  $U_A=\otimes_{i=1}^m u_{A_i} $ and $U_B=\otimes_{j=1}^n u_{B_j} $, to  both  $\rho$ and $\sigma$. Here the $u_{A_i}$ and $u_{B_j}$ are sampled from   a
		unitary 2-design with dimension $\ell$. Next, we perform projective measurements in the computational basis $|\mathbf{s}_{AB}\rangle =|\mathbf{s}_A\rangle\otimes |\mathbf{s}_B\rangle  =(\otimes_{i=1}^m|s_{A_i}\rangle) \otimes (\otimes_{j=1}^n|s_{B_j}\rangle )$, recording the outcomes $\mathbf{s}_{AB}$ and  $\mathbf{t}_{AB}$ for the two states respectively.  Repeating this measurement for the same $U_{AB}$ allows us to estimate:
		$
		\mathcal{P}(U_{X}, \mathbf{s}_{X}) = \tr[U_{X} \rho_X U_{X}^\dagger |\mathbf{s}_{X}\rangle  \langle \mathbf{s}_{X}| ], 
		\mathit{Q}(U_{X}, \mathbf{t}_{X}) = \tr[U_{X} \sigma_X U_{X}^\dagger |\mathbf{t}_{X}\rangle  \langle \mathbf{t}_{X} | ],
		$ 
		where $X\in \{A,B,AB\}$ and $\rho_{AB}=\rho,\sigma_{AB}=\sigma$. By repeating this procedure with varying $U_{AB}$, we can estimate the state overlaps $\tr[\rho_X \sigma_X]$ using second-order cross-correlations:
		$
		\tr[\rho_X \sigma_X]= d_X \sum_{\mathbf{s}_X,\mathbf{t}_X} (-\ell)^{-\mathcal{D}(\mathbf{s}_X,\mathbf{t}_X)} \mathbb{E}_{U_X} [\mathcal{P}(U_{X}, \mathbf{s}_{X}) \mathit{Q}(U_{X}, \mathbf{t}_{X})],
		$
		where $d_X$ denotes the dimension of system $X$, $\mathcal{D}(\mathbf{s}_X, \mathbf{t}_X)$ is the Hamming distance between $\mathbf{s}_X$ and $\mathbf{t}_X$, and $\mathbb{E}_{U_X}$ indicates the ensemble average over random unitaries of the form $U_X$. As $\langle \rho_X, \sigma_X\rangle= \tr[\rho_X\sigma_X]$, we can estimate $\mathcal{S}(\rho,\sigma) = \max\{\frac{\langle \rho, \sigma\rangle}{\langle \rho_A, \sigma_A\rangle},\frac{\langle \rho, \sigma\rangle}{\langle \rho_B, \sigma_B\rangle}\}$. Then we get the estimation of  $\mathrm{SN}(\rho),\mathrm{SN}(\sigma) \geq \lceil\mathcal{S}(\rho,\sigma) \rceil$ (see also in Eq. \eqref{eq:Schmidt_estimation}).  }\label{fig:inner_product_circuit} 
\end{figure}

(ii) The quantity \(\mathcal{S}(\rho, \sigma)\) can be determined by using the local randomized measurement techniques as demonstrated in Ref.~\cite{IPC_Elben20}, which allows us to estimate the Schmidt numbers of \(\rho\) and \(\sigma\). Note that the same  data used to estimate global state overlap can also be utilized to estimate local state overlap, see Fig.~\ref{fig:inner_product_circuit}. In addition, state overlap can be estimated by using the swap test, which involves the application of a Hadamard gate, followed by a controlled swap gate, another Hadamard gate and then the measurement on the ancillary qubit.

(iii) Theorem 1 implies that, in addition to their well-known role in characterizing entanglement, the purities of global and local states also provide insights into the Schmidt number. In fact, if the local R\'enyi entropy $S(\rho_X):= -\log_2 \tr[\rho_X^2]$ exceeds the global one, $S(\rho)=-\log_2 \tr[\rho^2]$ (where $X\in\{A,B\}$ denotes the local system), it indicates the presence of entanglement. However, the physical meaning of the difference between the global and local R\'enyi entropies, i.e., $ \Delta_X(\rho):=S(\rho_X)-S(\rho)$, has so far remained unclear.   According to Theorem 1, by choosing $\sigma=\rho$, we find that the magnitude of this difference directly characterizes the Schmidt number. Specifically, if $\Delta_X(\rho) >\log_2 r$, then $\mathrm{SN}(\rho)\geq r+1.$

(iv) The $r$-IPC provides a tight method to detect the Schmidt numbers of all pure states.
Let $\rho=|\Phi\rangle \langle\Phi|$  be a pure state of Schmidt number $r\geq 2$. Then there exists  a  state $ \hat{\rho}=|\hat{\Phi}\rangle\langle\hat{\Phi}|$ such that 
$\mathcal{S}(\rho,\hat{\rho})=r$. 
In fact, if $ |\Phi\rangle=\sum_{i=1}^r \sqrt{\lambda_i} |e_i f_i\rangle $ is its Schmidt decomposition,  then  $|\hat{\Phi}\rangle=\mathcal{N}\sum_{i=1}^r \frac{1}{\sqrt{\lambda_i}} |e_i f_i\rangle $ where $\mathcal{N}=1/(\sum_{i=1}^n \frac{1}{\lambda_i})$ is a normalized factor. From this point, to certify the Schmidt number of $|\Phi\rangle$ via the $r$-IPC method, the best approach is to choose $|\hat{\Phi}\rangle$ as its verifier. In this context, since $\mathcal{S}(\rho, \hat{\rho})$ is an integer (specifically the Schmidt number itself) and thus discrete, the $r$-IPC method for pure states exhibits robustness to noise.

	(v)  A typical method  used for Schmidt number detection  is the $r$-fidelity based criterion ($r$-FBC): for  a pure   state  $|\Phi\rangle$ in $ \mathcal{H} $ with $\mathrm{SN}(|\Phi\rangle)\geq r$  and  Schmidt decomposition $\sum_{k=1}^K  \sqrt{\lambda_k} |e_k f_k\rangle$ where  $\lambda_1\geq \lambda_2\geq  \cdots \geq \lambda_K>0$, define    
	\begin{equation}\label{eq:r_FBC}
		\mathcal{W}_{r-\mathrm{FBC}}(\Phi):= \left(\sum_{k=1}^r \lambda_k\right) \mathbb{I}_{AB}- |\Phi\rangle\langle \Phi|.	
	\end{equation} 
For any state $\rho \in \mathbb{D}(\mathcal{H})$ satisfying $\mathrm{SN}(\rho)\leq r$, one has
$
\mathrm{Tr}\!\left[\rho\,\mathcal{W}_{r\text{-}\mathrm{FBC}}(\Phi)\right]\geq 0 .
$
Consequently, the observation of
$\mathrm{Tr}[\rho\,\mathcal{W}_{r\text{-}\mathrm{FBC}}(\Phi)]<0$
certifies that the Schmidt number of $\rho$ is strictly larger than $r$. A state $\rho$ is called an \emph{$(r+1)$-unfaithful state} \emph{($(r+1)$-UFS)} if it is undetectable by the $r$-FBC, namely,
$
\mathrm{Tr}\!\left[\rho\,\mathcal{W}_{r\text{-}\mathrm{FBC}}(\Phi)\right]\geq 0
$
for all witnesses $\mathcal{W}_{r\text{-}\mathrm{FBC}}(\Phi)$ of the form given in
Eq.~\eqref{eq:r_FBC}.
States of this type demonstrate an intrinsic limitation of the $r$-FBC.
 Of particular interest are $(r+1)$-UFSs whose Schmidt numbers are strictly larger than $r$.
 We show that the $r$-IPC  can nevertheless detect a nontrivial subset of these states, thereby motivating the following definition.

A state $\rho$ is called an \emph{$(r+1)$-UFS-IP state} if it is an $(r+1)$-UFS and its Schmidt number can  be detected by the $r$-IPC, i.e., if there exists a state $\sigma$ such that
\begin{equation}
	\langle \rho,\sigma\rangle >
	\min\!\left\{
	r\,\langle \rho_A,\sigma_A\rangle,\,
	r\,\langle \rho_B,\sigma_B\rangle
	\right\}.
\end{equation}

Each $(r+1)$-UFS-IP state has Schmidt number strictly greater than $r$, can be certified by the $r$-IPC, but fails to be detected by the $r$-FBC. 
This separation highlights a fundamental operational distinction between the $r$-FBC and the $r$-IPC.
To characterize $(r+1)$-UFSs, it is therefore necessary to establish bounds on the detectability of the $r$-FBC. In particular, we show below that any state violating the $r$-FBC must satisfy a specific spectral bound condition.

	\emph{Proposition 1} [Spectrum bound of high dimensional entanglement via $r$-FBC]
	Let $\mathcal{H}=\mathcal{H}_A\otimes \mathcal{H}_B$ be system with local dimensions $d_A$ and $d_B$ and $\rho\in\mathbb{D}(\mathcal{H})$.  Let $\lambda(\rho)$ denote the largest eigenvalue of $\rho$. If 
	$\lambda(\rho)\leq \max\{\frac{r}{d_A}, \frac{r}{d_B}\}$, then all $r$-fidelity based witness $\mathcal{W}_{r-\mathrm{FBC}}(\Phi)$  fail to detect the Schmidt number of   $\rho $, i.e., $\tr[\rho\mathcal{W}_{r-\mathrm{FBC}}(\Phi) ]\geq 0.$ 
	
	The proof of Proposition 1 is provided in Appendix C.	
	This bound allows us to the following examples  showing that the $r$-IPC can be used to detect $(r+1)$-UFS,  thus outperforming the $r$-FBC (for more examples, see Appendix C).

	\emph{Example 2.---}($r$-IPC vs $r$-FBC) We consider a variant of isotropic states in $\mathcal{H}_{A}\otimes \mathcal{H}_{B}$ with $d_A=d_B=d\geq 3$
\begin{equation}\label{eq:example2}
	\rho(x)=(1-x)\frac{ \mathbb{I}_{(d-1)^2}}{(d-1)^2} + x|\Psi\rangle\langle \Psi|,
\end{equation}
where $\mathbb{I}_{(d-1)^2}= \sum_{i=0}^{d-2} \sum_{j=0}^{d-2} |ij\rangle \langle ij|$ and $|\Psi\rangle$ is the maximally entangled state  $\frac{1}{\sqrt{d}} \sum_{i=0}^{d-1} |ii\rangle$. If we select $\sigma(y)=|\Theta(y)\rangle\langle \Theta(y)|,$
where
$|\Theta(y)\rangle= \left(\sum_{i=0}^{d-2} y |ii\rangle\right) +\sqrt{1-(d-1)y^2}|(d-1)(d-1)\rangle,$ and find a $y_x\in [0, \frac{1}{\sqrt{d-1}}]$ such that $\mathcal{S}(\rho(x),\sigma(y_x))$ reaches its maximal value. Utilizing the above \(\sigma(y_x)\) and the spectral bound condition from Proposition 1, the \(r\)-IPC detected some \((r+1)\)-UFSs of the form \(\rho(x)\), which are plotted in Fig.~\ref{fig:Example2}. The detailed analyses are included in Appendix C.

	 \begin{figure}
		\centering
		\includegraphics[scale=0.685]{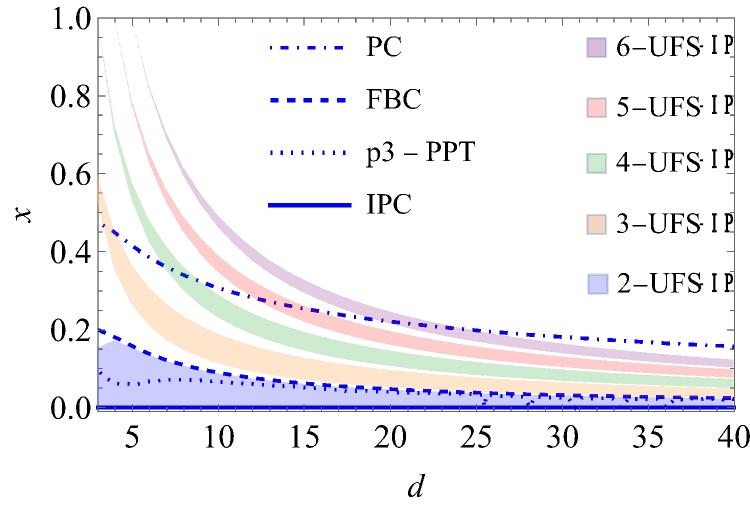}	
		\caption{The figure presents two complementary results for the state defined in Eq.~\eqref{eq:example2}, which is represented by a point $(d,x)$ in the parameter space. 
			Note that $\rho(x)$ is entangled for all $x>0$, and that larger values of $x$ correspond to a higher degree of entanglement in $\rho(x)$.			
			(a) The $(r+1)$-UFS-IP states are indicated by the five colored regions ($r=1,2,3,4,5$). 
			For each region, the lower boundary is determined by the saturation condition
			$\mathcal{S}(\rho(x_*),\sigma(y_{x_*})) = r$, 
			above which $\mathcal{S}(\rho(x),\sigma(y_x)) > r$ holds and the states can therefore be detected by the $r$-IPC. 
			The upper boundary of the corresponding region is given by the equality $\lambda(\rho(x)) = r/d$, 
			below which the states necessarily belong to the $(r+1)$-UFS by Proposition 1.			
			(b) The IPC exhibits superior detection performance compared to three experimentally accessible entanglement criteria, namely the $p_3$-PPT criterion, the FBC, and the purity criterion (PC). 
			This advantage is illustrated by the four blue curves: states lying above a given curve are detected as entangled by the corresponding method, and a lower curve therefore indicates a stronger entanglement-detection capability. }\label{fig:Example2}
	\end{figure}

	(vi)  Henceforth for $r=1$, we call $1$-IPC (resp. $1$-FBC) IPC (resp. FBC) for simplicity. The IPC can be utilized to detect quantum entanglement. Now we compare the performance of IPC with that of  three other  experimentally accessible entanglement criteria: purity criterion (PC), fidelity-based criterion (FBC) and  $p_3$-PPT criterion.     Let   $p_k= \mathrm{Tr}[(\rho ^{\mathrm{T}_A})^k]$ denote the $k$th moment of   $\rho ^{\mathrm{T}_A}$, where the $\mathrm{T}_A$ here denotes the partial transpose operator with respective to subsystem $A$ . Then the $p_3$-PPT   criterion says  that if   $\rho$ is separable,  then $p_2^2\leq p_3$.
	
	In this setting, we observe that the IPC provides a strictly advantage over both PC and FBC. It also outperforms the $p_3$-PPT criterion in specific instances. We compare the  performances of IPC, $p_3$-PPT, FBC, and PC on the states defined by Eq. \eqref{eq:example2}.    The results are illustrated by the four blue lines in Fig. \ref{fig:Example2}, with detailed calculations provided in Appendix D. Additionally, the IPC approach can be extended to detect entanglement in multipartite systems (see Appendix E).

	\vskip 8pt

	\emph{Relations between $r$-$\mathrm{IPC}$ and    other terms in  quantum information theory.---} We find that the $r$-IPC is connected to some other significant concepts in quantum information theory, such as distillation and teleportation. It is well established that for system with a global dimension greater than 6, there exist bound entangled states,  i.e.,  entangled states that cannot be distilled into maximally entangled states using local operations and classical communication (LOCC).  It has been shown  that all entangled states that violate the reduction criterion are distillable \cite{CC_Reduction_HH_99}. Consequently, we can infer that all entangled states detectable by the 1-IPC are also distillable.

	Note that the Schmidt number remains invariant under local unitary operations, i.e., $\mathrm{SN}(U\otimes V \rho U^\dagger\otimes V^\dagger)=\mathrm{SN}( \rho )$ for all unitary matrices $U$ and $ V$ acting on systems 
	$A$ and $B,$ respectively. However, this is not the case for the ${\mathcal{S}}(\rho,\sigma)$. Hence, we are motivated to define
	\begin{equation}\label{eq:Sch_Estimation_Max}
		\hat{\mathcal{S}}(\rho,\sigma)=\max_{U, V}\{\mathcal{S} ( U\otimes V \rho U^\dagger\otimes V^\dagger,\sigma) \},
	\end{equation} 
	where the maximization is taken over all unitary matrices $U$ and $V$ acting on systems 
	$A$ and $B,$ respectively.  We also have  
	$
	\mathrm{SN}({\rho})\geq \lceil  \hat{\mathcal{S}}(\rho,\sigma) \rceil, $ and  $\mathrm{SN}({\sigma})\geq \lceil  \hat{\mathcal{S}}(\rho,\sigma) \rceil$  which provides an improved lower bound compared to the one given in   \eqref{eq:Schmidt_estimation}.
	If $d_A = d_B = d$ and $\sigma$ is chosen to be the maximally entangled state $|\Psi\rangle\langle \Psi|$ where $|\Psi\rangle := \frac{1}{\sqrt{d}}\sum_{i=0}^{d-1} |ii\rangle$, then by  simple algebraic calculations, we obtain
	\begin{equation}\label{eq:Schmidt_Tele}
		\hat{\mathcal{S}}(\rho,|\Psi\rangle\langle \Psi|) = d\mathrm{F}_\rho,
	\end{equation}
where $\mathrm{F}_\rho := \max_{U \in \mathrm{U}(d)} \langle \Psi | (\mathbb{I}_d \otimes U^\dagger) \rho (\mathbb{I}_d \otimes U) |\Psi\rangle$ 	is the fully entangled fraction (also known as maximal
singlet fraction)   that plays a crucial role in quantifying entanglement and the feasibility of teleportation \cite{FEF1,FEF2}. Formula \eqref{eq:Schmidt_Tele} clearly illustrates the connection between $\hat{\mathcal{S}}(\rho,|\Psi\rangle\langle \Psi|)$ and the fully entangled fraction $\mathrm{F}_\rho$, emphasizing the significance of the measure $\hat{\mathcal{S}}(\rho,\sigma)$  in the field of quantum information. Deriving a closed-form analytical expression for $\hat{\mathcal{S}}(\rho, \sigma)$ is  theoretically  a challenging task. However,  $\hat{\mathcal{S}}(\rho, \sigma)$ can be evaluated using variational quantum algorithms. To be more explicit, this corresponds to tackling the following  optimization problem:
	\begin{equation}\label{eq:Sch_VQE}
		\sup_{\boldsymbol{\theta}, \boldsymbol{\xi}}\left\{\mathcal{S}\left( U(\boldsymbol{\theta})\otimes V(\boldsymbol{\xi}) \rho U(\boldsymbol{\theta})^\dagger\otimes V(\boldsymbol{\xi})^\dagger,\sigma \right) \right\},
	\end{equation}
	where $U(\boldsymbol{\theta})$ and $V(\boldsymbol{\xi})$ represent parameterized unitary operators. This topic is beyond the scope of the current work and will be the subject of dedicated investigation in future research.

\vskip 8pt

\emph{Conclusion and Discussions.---} In this work, we have proposed an approach for detecting the Schmidt number by leveraging the ratio of global to local state overlaps between two quantum states. One of the most unique aspects of our method, compared to common approaches, is that it simultaneously probes the entanglement properties of two  unknown quantum states. We have demonstrated several key advantages of this approach, including its practical implementation and  the effectiveness compared with  other entanglement detection techniques. By utilizing the state overlaps, which can be efficiently measured through local randomized measurements or the swap test, our approach presents an experimentally feasible illustration of the well-established $r$-reduction criterion. The $r$-IPC can be used to detect certain $(r+1)$-unfaithful states and can precisely determine the Schmidt number of all pure states with  robustness. We have shown that the IPC is strictly stronger than  both purity and fidelity-based criteria and even outperform the $p_3$-PPT criterion by detailed examples.  These results contribute to deepening our understanding of entanglement detection methodologies and highlight advancements in the physical implementation.

Several interesting problems remain to be addressed, including: (1) whether  IPC is always  stronger than $p_3$-PPT criterion,  (2) whether $r$-IPC is more powerful than $r$-FBC for $r \geq 2$,  and  (3)  how to extend the IPC approach to detect genuine multipartite entanglement. The global and local   purity \cite{CC_Pure_HH96}, Rényi entropy \cite{EX_RM__Renyi_Brydges19},  and  disorder \cite{CC_TM_Major_Nielsen01}   of one state have been used to  identify quantum entanglement.  In this work, we  have taken into account the global and local inner products to certify entanglement of a pair of states.   Similar properties may also be extended to quantities, such as trace norm,   von Neumann entropy, and   other distance based quantities. According to (iv), the $r$-IPC may offer certain advantages in verifying the Schmidt number of pure states. We look forward to an experimental validation in this regard.

\vskip 20pt

\begin{acknowledgements}
This work is supported by National Natural Science Foundation of China (Grant Nos. 12371458, 12371132, 12305007, 12501639, 62072119, 12075159, 12171044, 12071179, 
 12075159 and 12171044), 
the Fundamental Research Funds for the Central Universities, the Guangdong Basic and Applied Basic Research Foundation under Grants Nos. 2024A1515010380, 2024A1515030023, Key Lab of Guangzhou for Quantum Precision Measurement under Grant No.202201000010,   Anhui Provincial
Natural Science Foundation (Grant No. 2308085QA29);
the Alexander von Humboldt Foundation, Natural Science Foundation of Shanghai (Grant No. 20ZR1426400),
Shenzhen Institute for Quantum Science and Engineering, Southern University of Science and Technology
(Grant Nos. SIQSE202005),  the Academician Innovation Platform of Hainan Province, the
Deutsche Forschungsgemeinschaft (DFG, German Research
Foundation, project number 563437167), the Sino-German
Center for Research Promotion (Project M-0294), the German
Federal Ministry of Research, Technology and Space (Project
QuKuK, Grant No. 16KIS1618K and Project BeRyQC, Grant
No. 13N17292).
\end{acknowledgements}

\newpage

 \onecolumngrid
 \appendix

\section{Appendix A: Proof of Theorem 1}
\emph{Theorem 1.---}(Inner product criterion for Schmidt number) 	Let $\rho\in \mathbb{D}(\mathcal{H}_A\otimes\mathcal{H}_B)$ be   a bipartite density matrix. If  $\mathrm{SN}(\rho) \leq r$, then for all  $\sigma\in \mathbb{D}(\mathcal{H}_A\otimes\mathcal{H}_B)$, we have  
	\begin{equation}\label{eq:Inner_ineq_SCHM}
		\langle \rho,\sigma\rangle\leq \min\{ r \ \langle \rho_A,\sigma_A\rangle,r\ \langle \rho_B,\sigma_B\rangle\}.
	\end{equation}

\begin{proof}	If  $\mathrm{SN}(\rho) \leq r$ and $\sigma$ is any state in $\mathbb{D}(\mathcal{H}_A\otimes\mathcal{H}_B)$,  then $\rho$ and $\sigma$ can be written as 
	\begin{equation}
		\begin{array}{rcl}
			\rho&=&\displaystyle\sum_{i=1}^mp_i \rho_i=\sum_{i=1}^mp_i |\psi^{AB}_i\rangle  \langle\psi^{AB}_i |,\\[2mm]   
			\sigma&=&\displaystyle\sum_{j=1}^n q_j \sigma_j=\sum_{j=1}^n q_j|\phi_j^{AB}\rangle\langle \phi_j^{AB}|,
		\end{array} 
	\end{equation}  
	where $p_i,q_j\geq 0$, $\mathrm{SN}(|\Psi_i^{AB}\rangle ) \leq r$ and  $\sigma$ is given by its   spectrum decomposition.
	By the $\mathbb{R}-$linearity of inner product and partial trace, we have 
	\begin{align}
		r\langle \rho_X,\sigma_X\rangle-\langle \rho,\sigma\rangle=\sum_{i=1}^m\sum_{j=1}^n  p_iq_j\left(r\langle {(\rho_i)}_X,{(\sigma_j)}_X\rangle-\langle \rho_i,\sigma_j\rangle\right).
	\end{align}
	It is  sufficient to prove that $r\langle {(\rho_i)}_X,{(\sigma_j)}_X\rangle-\langle \rho_i,\sigma_j\rangle\geq 0 $  for each $i,j.$	 Therefore, we can assume both $\rho$ and $\sigma$ are pure states. Without loss of generality, we  suppose that $\rho$ and $\sigma$  are density matrices of the pure states  $\sum_{k=1}^K \sqrt{\lambda_k}|e_k\rangle_A|f_k\rangle_B $ (here $K\leq r$ and  $\lambda_k>0$ for all $1\leq k\leq K$) and $\sum_{i=1}^{d_A}\sum_{j=1}^{d_B} a_{ij}|e_i f_j\rangle_{AB}$   (here $\{|e_i\rangle_A\}_{i=1}^{d_A} $ and $\{|f_j\rangle_B\}_{j=1}^{d_B} $ are some other orthonormal basis of subsystem $A$ and $B$). In this case $\langle \rho,\sigma\rangle=|\sum_{k=1}^K \sqrt{ \lambda_k} a_{kk}|^2.$ Moreover, we have 
    \begin{eqnarray}
		\displaystyle 	\rho_A&=&\displaystyle \sum_{k=1}^K \lambda_k  |e_k\rangle_A\langle e_k|, \quad  \sigma_A=\displaystyle \sum_{i,k=1}^{d_A} \left(\sum_{l=1}^{d_B}a_{il}a_{kl}^\dagger\right) |e_i\rangle_A\langle e_k|, 
        \nonumber
        \\
		\displaystyle	\rho_B&=&\displaystyle \sum_{k=1}^K \lambda_k  |f_k\rangle_B\langle f_k|,\quad  \sigma_B=\displaystyle \sum_{j,l=1}^{d_B} \left(\sum_{i=1}^{d_A}a_{ij}a_{il}^\dagger\right) |f_j\rangle_B\langle f_l|.
	\end{eqnarray}
	Therefore, $\langle \rho_A,\sigma_A\rangle= \sum_{k=1}^K(\lambda_k \sum_{l=1}^{d_B}|a_{kl}|^2)$ and  $\langle \rho_B,\sigma_B\rangle= \sum_{k=1}^K(\lambda_k \sum_{i=1}^{d_A}|a_{ik}|^2).$ By the Cauchy-Schwarz inequality, we have
	$$\langle \rho,\sigma\rangle=\left|\sum_{k=1}^K  \sqrt{\lambda_k} a_{kk}\right|^2\leq  \left(\sum_{k=1}^K   \lambda_k| a_{kk}|^2\right)\sum_{k=1}^K 1\leq K  \left(\sum_{k=1}^K \lambda_k| a_{kk}|^2\right)\leq  \min\{K\langle \rho_A,\sigma_A\rangle,K\langle \rho_B,\sigma_B\rangle\}\leq \min\{r\langle \rho_A,\sigma_A\rangle,r\langle \rho_B,\sigma_B\rangle\} .  $$
	This completes the proof.
\end{proof}

\section{Appendix B: Proof of Observation 1}

In order to prove Observation 1, we need the following two lemmas, which can be verified by direct inspection.

\emph{Lemma 1.---}Let $P, Q\in \mathbb{L}(\mathcal{H}_A\otimes \mathcal{H}_B)$ and denote 
$P_X:=\tr_{\overline{X}}[P],$ $ Q_X:=\tr_{\overline{X}}[Q]$ where $X\in\{A,B\}$ and $\overline{X}:=\{A,B\}\setminus X$. Then  the following equalities always hold
\begin{equation}\label{eq:global_local}
	\langle \mathbb{I}_A\otimes P_B,\   Q \rangle=
	\langle  P_B,\  Q_B\rangle, \text{ and } \langle P_A\otimes \mathbb{I}_B,\   Q \rangle=\langle  P_A,\  Q_A\rangle.
\end{equation}

\kommentar{
\begin{proof} Here we just prove the first one as the second one follows similarly. Fixed $P$, the operator
	$L_P:=\langle \mathbb{I}_A\otimes P_B,\   \cdot \rangle-
	\langle  P_B,\  {(\cdot) }_B\rangle$ defined by sending $Q$ to $\langle \mathbb{I}_A\otimes P_B,\   Q \rangle-
	\langle  P_B,\  Q_B\rangle$ can be looked as  a linear operation from the linear space $\mathbb{L}(\mathcal{H}_A\otimes \mathcal{H}_B)$ to $\mathbb{C}$. We need to prove that $L_P$ is a zero operation. It is suffcient to show that 
	$L_P$ acts trivially (zero) on a basis of $\mathbb{L}(\mathcal{H}_A\otimes \mathcal{H}_B)$.
	In fact, $\{|i\rangle \langle j| \otimes |k\rangle \langle l|\}$  forms a basis.
	Clearly,
	\begin{equation}
		\begin{array}{ cl}
			&\displaystyle \langle \mathbb{I}_A\otimes P_B,\  |i\rangle \langle j| \otimes |k\rangle \langle l| \rangle-
			\langle  P_B,\  (|i\rangle \langle j| \otimes |k\rangle \langle l|)_B\rangle \\[2mm]
			=& \displaystyle \tr [|i\rangle \langle j| \otimes P_B (|k\rangle \langle l|)]-\tr[P_B (|k\rangle \langle l|) ] \delta_{ij}\\[2mm]
			=& \tr [|i\rangle \langle j|] \cdot  \tr[P_B (|k\rangle \langle l|)]-\tr[P_B (|k\rangle \langle l|) ] \delta_{ij}=0.
		\end{array}	     	
	\end{equation} 
\end{proof}
} %

\emph{Lemma 2.---}Let $P$ be any Hermitian operation on $\mathcal{H}$. Define a map $f_P: \mathbb{D}(\mathcal{H}) \rightarrow \mathbb{R}$ by sending $Q$ to $\tr[PQ]$. Then $P$ is positive semidefinite if and only if 
$f_P$  takes  nonnegative  values on each $Q$, i.e.,
$f_P(Q) \geq 0$ for any $Q\in \mathbb{D}(\mathcal{H}).$
\kommentar{
\begin{proof}
	If $P$ is positive semidefinite, then for any $|\psi\rangle \in \mathcal{H}$, $\langle \psi|P|\psi\rangle \geq 0.$ As any $Q\in \mathbb{D}(\mathcal{H})$ is also  positive semidefinite, by spectrum decomposition, we have
	$$ Q=\sum_{i} \lambda_i |\psi_i\rangle \langle \psi_i |$$
	where $\lambda_i\geq 0$.	  Therefore,
	$$f_P(Q)=\tr[PQ]=\sum_i\lambda_i \tr[P |\psi_i\rangle \langle \psi_i |]= \sum_i\lambda_i \langle \psi_i|P|\psi_i\rangle \geq 0.$$
	
	If $P$ is not positive semidefinite, then it admits some negative eigenvalue $\lambda<0$. Suppose 
	$ P|\psi\rangle =\lambda |\psi\rangle$, where $|\psi\rangle$ is a unit vector in $\mathcal{H}$. Denote $Q=|\psi\rangle\langle \psi|$. We have 
	$$ f_P(Q)=\tr[PQ]=\tr[P|\psi\rangle \langle \psi|]=\lambda\tr[|\psi\rangle \langle \psi|]=\lambda<0.$$
\end{proof} 
} %

We can then proceed with the proof of the Observation 1:

	\emph{Observation 1.---}(Equivalence of $r$-IPC and $r$-RC) A bipartite state $\rho\in \mathbb{D}(\mathcal{H}_A\otimes\mathcal{H}_B)$ violates the inequality \eqref{eq:Inner_ineq_SCHM} for some state $\sigma$ if and only if it violates the inequalities \eqref{eq:r_RC}. 
    
\begin{proof}
	
	By Lemma 2, the state $\rho$ violates inequality  \eqref{eq:r_RC} if and only if  
	at least one of $f_{P_1}$ and $f_{P_2}$  assumes negative value for some state $\sigma$, here
	$P_1:=r\mathbb{I}_A \otimes \rho_B -\rho $ and $P_2:=r \rho_A \otimes \mathbb{I}_B  -\rho .$ That is,   $ f_{P_1}(\sigma)<0$ or $ f_{P_2}(\sigma)<0$. By Lemma 1, we have
	\begin{align}
		f_{P_1}(\sigma)=\langle  P_1, \sigma \rangle =r \langle  \mathbb{I}_A \otimes \rho_B   , \sigma \rangle -\langle \rho, \sigma\rangle=r   \langle    \rho_B   , \sigma_B \rangle -\langle \rho, \sigma\rangle,\nonumber\\[2mm] 
		f_{P_2}(\sigma)=\langle  P_2, \sigma \rangle = r \langle  \rho_A \otimes \mathbb{I}_B   , \sigma \rangle -\langle \rho, \sigma\rangle= r  \langle    \rho_A   , \sigma_A \rangle -\langle \rho, \sigma\rangle.
	\end{align}
	Hence the statement  $ f_{P_1}(\sigma)<0$ or $ f_{P_2}(\sigma)<0$  for some $\sigma$ is equivalent to 
	$  \min\{r\langle \rho_A,\sigma_A\rangle,r\langle \rho_B,\sigma_B\rangle\}< \langle \rho,\sigma\rangle$,
	which implies the violation of inequality of \eqref{eq:Inner_ineq_SCHM} for this $\sigma.$ 
\end{proof}

\section{Appendix C:   $r$-IPC witness $r$-unfaithful entanglement}

To describe the $(r+1)$-unfaithful states,  we first establish certain bounds on the detectability of the $r$-FBC  in the following proposition.

\vskip 5pt

\emph{Proposition  1} [Spectrum bound of high dimensional entanglement via $r$-FBC]
Let $\mathcal{H}=\mathcal{H}_A\otimes \mathcal{H}_B$ be system with local dimensions $d_A$ and $d_B$ and $\rho\in\mathbb{D}(\mathcal{H})$.  Let $\lambda(\rho)$ denote the largest eigenvalue of $\rho$. If 
$\lambda(\rho)\leq \max\{\frac{r}{d_A}, \frac{r}{d_B}\}$, then all $r$-fidelity based witness $\mathcal{W}_{r-\mathrm{FBC}}(\Phi)$  fail to detect the Schmidt number of   $\rho $, i.e., $\tr[\rho\mathcal{W}_{r-\mathrm{FBC}}(\Phi) ]\geq 0.$     

\begin{proof}
	Suppose that $\rho$ can be witnessed by the detector $\mathcal{W}_{r-\mathrm{FBC}}(\Phi) =\left(\sum_{k=1}^r \lambda_k\right)\mathbb{I}_{AB} -|\Phi\rangle\langle \Phi|,$ with
	$|\Phi\rangle$  is an entangled pure state,   $|\Phi\rangle=\sum_{k=1}^l  \sqrt{\lambda_k} |e_k f_k\rangle$ where $l\geq r$, $\{|e_k\rangle\}_{k=1}^l$ and $\{|f_k\rangle\}_{k=1}^l$ being  pairwise orthonormal vectors of subsystems $A$ and $B$ respectively, and $\lambda_1\geq \lambda_2\geq \cdots \geq \lambda_l>0$. To witness the  Schmidt number of $\rho$ via $\mathcal{W}_{r-\mathrm{FBC}}(\Phi) $, one has 
	$\tr[\rho\mathcal{W}_{r-\mathrm{FBC}}(\Phi) ] <0$. That is,
	\begin{equation} \sum_{k=1}^r \lambda_k< \langle \Phi| \rho |\Phi\rangle.
	\end{equation}
	On the other hand, since $\lambda(\rho)$ is the largest eigenvalue of $\rho$,  
	$\lambda(\rho) \mathbb{I}_{AB} -\rho$ is  a positive semidefinite operator.   Hence, 
	$\langle \Phi | \lambda(\rho) \mathbb{I}_{AB} -\rho| \Phi\rangle \geq 0$, which implies that 
	$ \lambda (\rho)\geq \langle \Phi| \rho |\Phi\rangle.$ Therefore, we have $\lambda(\rho)>\sum_{k=1}^r \lambda_k$. 
	As $\lambda_1\geq \lambda_2\geq \cdots \geq \lambda_l>0$, the following inequality holds 
   \begin{equation}1=\lambda_1+\lambda_2+\cdots +\lambda_l\leq \sum_{k=1}^r \lambda_k+\frac{l-r}{r}\sum_{k=1}^r \lambda_k=\frac{l}{r}\sum_{k=1}^r \lambda_k\leq \frac{\min\{d_A,d_B\}}{r}\sum_{k=1}^r \lambda_k.\end{equation}
	We deduce that $\lambda(\rho)>\sum_{k=1}^r \lambda_k\geq  \frac{r}{\min\{d_A,d_B\}} =\max\{\frac{r}{d_A},\frac{r}{d_B}\}$ which contradict with our assumption.
\end{proof}

 As a consequence, if $\lambda(\rho)\leq \max\{\frac{r}{d_A}, \frac{r}{d_B}\}$  holds, the state $\rho$ must be    $r$-unfaithful. 
Now based on the above proposition, we  could present  some examples to show that $r$-IPC outperforms the $r$-FBC. That is, how to find states that are $(r+1)$-UFS-IP.  First, we come back to the Example 2.
 In Example 2, we consider the following state, 
\begin{equation}
	\rho(x)=(1-x)\frac{ \mathbb{I}_{(d-1)^2}}{(d-1)^2} + x|\Psi\rangle\langle \Psi|
\end{equation}
where $\mathbb{I}_{(d-1)^2}= \sum_{i=0}^{d-2} \sum_{j=0}^{d-2} |ij\rangle \langle ij|$ and $|\Psi\rangle$ is the maximally entangled state  $\frac{1}{\sqrt{d}} \sum_{i=0}^{d-1} |ii\rangle$. 
 In order to show that the $r$-FBC fails  to detect the Schmidt numbers of some states of the form $\rho(x)$, we need to find out its maximal eigenvalue.  We will show  that  the maximal eigenvalue of  $\rho(x)$  is (see the derivation below),
 \begin{equation} \Delta:=\frac{m+nd+\sqrt{(m+nd)^2-4mn}}{2}\end{equation}
 where $m=\frac{1-x}{(d-1)^2},  n=\frac{x}{d}.$ 
 Therefore, by Proposition 1,  every  $r$-FBCs fail to detect the Schmidt number of $\rho(x)$ in Example 2 if $\Delta\leq \frac{r}{d}$, that is, 
 \begin{equation}\label{eq:r-FBC_I}
 	\Delta=\frac{m+nd+\sqrt{(m+nd)^2-4mn}}{2}   \leq \frac{r}{d}. 
 \end{equation}

Now we show how to use $r$-IPC to detect Schmidt number of $\rho(x)$.  We select $\sigma(y)=|\Theta(y)\rangle\langle \Theta(y)|,$
where
$|\Theta(y)\rangle= \left(\sum_{i=0}^{d-2} y |ii\rangle\right) +\sqrt{1-(d-1)y^2}|(d-1)(d-1)\rangle,$ and find  a $y=f(x)\in [0, \frac{1}{\sqrt{d-1}}]$ such that $\mathcal{S}(\rho(x),\sigma(y))$ reach  its  maximal value. Then we pink out the points $(x_r,d)$ where $\mathcal{S}(\rho(x_r),\sigma(f(x_r)))=r$ ($r=1,2,3,4,5$) for each $d$. Utilizing these numerical points $(x_r,d)$ and the spectral bound condition in Proposition 1, the \(r\)-IPC detected some \((r+1)\)-unfaithful states of the form \(\rho(x)\), which are plotted in Fig.~\ref{fig:Example2}.

Now we are the time  to show that  the maximal eigenvalue of  $\rho(x)$  is 
$  \Delta=\frac{m+nd+\sqrt{(m+nd)^2-4mn}}{2}.$   In fact, we need to find the optimal state $|\varphi\rangle=\sum_{i,j=1}^d a_{ij}|ij\rangle$ such that 
\begin{equation}
\langle \varphi| \rho(x) |\varphi\rangle=\frac{1-x}{(d-1)^2}\sum_{i,j=1}^{d-1} |a_{ij}|^2 +\frac{ x}{d} \big|\sum_{i=1}^d  a_{ii}\big|^2
\end{equation}
assume a maximal value under the condition $\sum_{i,j=1}^d |a_{ij}|^2=1$.  Without loss of generality, we can assume that all $a_{ij}\geq 0$, otherwise we replace $a_{ij}$ by $|a_{ij}|$  without decreasing the target number. Moreover, we can assume that $a_{ij}=0$ for all $i\neq j$, otherwise we can replace $a_{ii}$ to be $\sqrt{a_{ii}^2+a_{ij}^2}$ with an increasing of the target number. Therefore, it is equivalent to 
maximize $\frac{1-x}{(d-1)^2}\sum_{i=1}^{d-1} a_{ii}^2 +\frac{x}{d} \big(\sum_{i=1}^d  a_{ii} \big)^2$
under the condition $\sum_{i=1}^d a_{ii}^2=1.$ Set $m=\frac{1-x}{(d-1)^2}, $ and $ n=\frac{x}{d}.$ This is equivalent to finding an optimal value of the quadratic form with matrix $P$,
\begin{equation}
P= \left[\begin{array}{cccccc}
	m+n & n &n& \cdots& n&n\\
	n& m+n& n& \cdots& n&n\\
	n&n& m+n& \cdots &n &n\\
	\vdots&\vdots&\vdots&\ddots&\vdots\\
	n&n&n& \cdots & m+n&n\\
	n&n&n& \cdots & n&n 
\end{array}\right].
\end{equation}
To calculate   the eigenvalues of $P$, we need to find out the characeristic polynomial
$\mathrm{Det}[\lambda \mathbb{I}_d- P].$ This is:

\begin{eqnarray}
	\mathrm{Det}[\lambda \mathbb{I}_d- P]&=& \left|\begin{array}{cccccc}
		\lambda -(m+n) & -n &-n& \cdots& -n&-n\\
		-n& \lambda -(m+n)& -n& \cdots& -n&-n\\
		-n&-n& \lambda -(m+n)& \cdots &-n &-n\\
		\vdots&\vdots&\vdots&\ddots&\vdots\\
		-n&-n&-n& \cdots & \lambda -(m+n)&-n\\
		-n&-n&-n& \cdots & -n&\lambda -n 
	\end{array}\right |
     \nonumber
     \\[8mm]
	&=& \left|\begin{array}{cccccc}
		\lambda -(m+n) & -n &-n& \cdots& -n&-n+0\\
		-n& \lambda -(m+n)& -n& \cdots& -n&-n+0\\
		-n&-n& \lambda -(m+n)& \cdots &-n &-n+0\\
		\vdots&\vdots&\vdots&\ddots&\vdots\\
		-n&-n&-n& \cdots & \lambda -(m+n)&-n+0\\
		-n&-n&-n& \cdots & -n&\lambda -(m+n)+m 
	\end{array}\right |
      \nonumber
      \end{eqnarray}
	\begin{eqnarray}
	\phantom{\mathrm{Det}[\lambda \mathbb{I}_d- P]}
    &=& \left|\begin{array}{cccccc}
		\lambda -(m+n) & -n &-n& \cdots& -n&-n \\
		-n& \lambda -(m+n)& -n& \cdots& -n&-n \\
		-n&-n& \lambda -(m+n)& \cdots &-n &-n \\
		\vdots&\vdots&\vdots&\ddots&\vdots\\
		-n&-n&-n& \cdots & \lambda -(m+n)&-n \\
		-n&-n&-n& \cdots & -n&\lambda -(m+n)  
	\end{array}\right |    \nonumber \\[8mm]
	&+ &  \left|\begin{array}{cccccc}
		\lambda -(m+n) & -n &-n& \cdots& -n&0 \\
		-n& \lambda -(m+n)& -n& \cdots& -n&0 \\
		-n&-n& \lambda -(m+n)& \cdots &-n &0 \\
		\vdots&\vdots&\vdots&\ddots&\vdots\\
		-n&-n&-n& \cdots & \lambda -(m+n)&0 \\
		-n&-n&-n& \cdots & -n&m  
	\end{array}\right |   \nonumber  \\[4mm]
	&=& \mathrm{Det}[(\lambda-m) \mathbb{I}_d -n   O(d)]+m    \mathrm{Det}[(\lambda-m) \mathbb{I}_{d-1} -n * O(d-1)]  \nonumber
	\\[4mm]
	&=& (\lambda-m)^{d-1}(\lambda-m -nd)+ m(\lambda-m)^{d-2}(\lambda-m -n(d-1))
	  \nonumber \\[8mm]
	&=& (\lambda-m)^{d-2}[(\lambda-m)(\lambda-m -nd) +m\lambda-m^2 -mn(d-1)]
      \nonumber \\  [4mm]	 
	&=& (\lambda-m)^{d-2}[(\lambda-m)(\lambda-m -nd) +\lambda-m -n(d-1)]  \nonumber \\ [4mm]
	&=& (\lambda-m)^{d-2}[\lambda^2-(m+nd) \lambda + mn], 
\end{eqnarray}  %
where $O(k)$ denotes the $k\times k$ matrix with all entries ones.	Therefore, the maximal eigenvalue of $P$ is 
\begin{equation}
\Delta:=\frac{m+nd+\sqrt{(m+nd)^2-4mn}}{2}.
\end{equation}

\begin{figure}[t]
	\centering
	\includegraphics[scale=0.8]{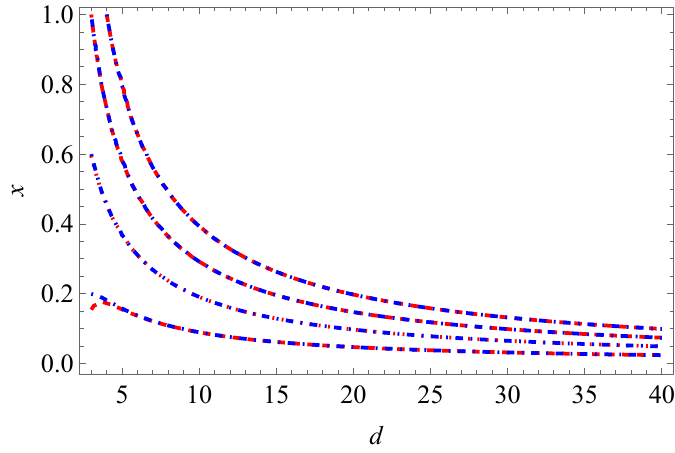}	
	\caption{This figure shows that the boundary obtained in Proposition~1 is  almost tight. 
		Specifically, we plot the two boundaries corresponding to Eqs.~\eqref{eq:r-FBC_I} and \eqref{eq:r-FBC_R} for $r=1,2,3,4$, ordered from bottom to top. 
		The blue curves correspond to the bound given by Eq.~\eqref{eq:r-FBC_R}, namely
		$
		x+\frac{1-x}{d(d-1)}=\frac{r}{d},
	$	above which the states can be detected by the $r$-FBC via the particular witness $\mathcal{W}_{r\text{-}\mathrm{FBC}}(\Psi)$. 
		The red curves correspond to the bound specified by Eq.~\eqref{eq:r-FBC_I}, i.e.,
	$
		\Delta=\frac{r}{d},
		$
			below which the states cannot be detected by the $r$-FBC. 
	 }\label{fig:r_FBC_IR} 
\end{figure}

\emph{Example 3.---}(2-IPC vs 2-FBC) The state $\rho_4=\frac{1}{2}|\Psi_3\rangle\langle \Psi_3\rangle +\frac{1}{2}|\Phi\rangle\langle \Phi|$, where $|\Psi_3\rangle=(|00\rangle+|11\rangle+|22\rangle| )/\sqrt{3}$ and $|\Phi\rangle= |23\rangle+|32\rangle)/\sqrt{2}$, is faithful but 3-unfaithful in system $\mathbb{C}^4\otimes \mathbb{C}^4$. 

If we select $\sigma(x)=|\Theta(x)\rangle\langle \Theta(x)|$, where
$|\Theta(x)\rangle=\sqrt{\frac{1}{3}+x}|00\rangle+\sqrt{\frac{1}{3}+x}|11\rangle +\sqrt{\frac{1}{3}-2x}|22\rangle$, we have $\mathcal{S}(\rho_4, \sigma(x))>2$ for all $0<x<\frac{1}{6}$, which attains the maximum value $\frac{12}{5}$ at $x=\frac{7}{54}.$ Hence $\mathrm{SN}({\rho_4})\geq 3$, and given the expression of $\rho_4$ we have $\mathrm{SN}({\rho_4})= 3.$  However, by the  Proposition 1, we obtain that $\rho_4$ could not be detected by $2$-FBC. Therefore, $\rho_4$ is 3-unfaithful.

\vskip 10pt

Now we illustrate that the bound in  Proposition 1 appears to be a strong bound.  On one hand,
  every  $r$-FBCs fail to detect the Schmidt number of $\rho(x)$ in Example 2 if $\Delta\leq \frac{r}{d}$, that is, 
\begin{equation}\label{eq:r-FBC_I}
	\Delta=\frac{m+nd+\sqrt{(m+nd)^2-4mn}}{2}   \leq \frac{r}{d} 
\end{equation}
where $m=\frac{1-x}{(d-1)^2}$ and $n=\frac{x}{d}$.  On the other hand, the particular witness  $\mathcal{W}_{r-\mathrm{FBC}}(\Psi)$ fails to detect the Schmidt number of $\rho(x)$ if and only if 
\begin{equation}\label{eq:r-FBC_R}
	x+\frac{1-x}{d(d-1)}\leq \frac{r}{d} . 
\end{equation} 
Comparing the two bounds in Eqs. \eqref{eq:r-FBC_I} and \eqref{eq:r-FBC_R}, it can be observed that the bound defined by \eqref{eq:r-FBC_I}   of $\rho(x)$ in Example 2 are nearly attained   by the boundary of the detectability of $\mathcal{W}_{r-\mathrm{FBC}}(\Psi)$ (see Fig. \ref{fig:r_FBC_IR} for an intuition).

\section{Appendix D: Comparison of IPC and some other entanglement criteria}

There are three experimentally accessible entanglement criteria, the purity criterion (PC), fidelity-based criterion (FBC), and $p_3$-PPT criterion which is a partial transpose (PT) moment-based criterion.   Now we compare the entanglement detectability of PC, FBC and  $p_3$-PPT criteria with that of IPC.

Denote $\mathcal{E}(\mathcal{H}_A\otimes \mathcal{H}_B) $ (or simply $\mathcal{E}$ for the known underlying systems) the set of all entangled states in the bipartite system  $\mathcal{H}_A\otimes \mathcal{H}_B$. Below we denote $\mathcal{E}_{\mathrm{Name}}$ the set of all entanglement states that can be detected by the criterion `$\mathrm{Name}$'. For example, we denote $\mathcal{E}_{\mathrm{IPC}}$ the set of entangled states that can be detected by the inner product criterion $\mathrm{IPC}$.

First, if we set $r=1$ and $\sigma=\rho$ in Theorem 1, we arrive at the purity criterion. Therefore, the IPC is stronger than the PC. Now we show that the IPC also outperforms the FBC.

\vskip 5pt

 \emph{Proposition 1.---}(IPC is stronger than FBC)   Any entangled state that could be detected by FBC can be also detected by IPC, i.e., 
$\mathcal{E}_{\mathrm{FBC}}\subseteq \mathcal{E}_{\mathrm{IPC}}$. 
\begin{proof}
	Suppose $|\Psi\rangle=\sum_{i=1}^l \sqrt{\lambda_i} |e_if_i\rangle$, where $\lambda_1\geq \lambda_2\geq\cdots\geq \lambda_l>0$ and  $\tr[\rho\mathcal{W}_{\Psi}]<0$. By the definition of $\mathcal{W}_{\Psi}$, we have
	$\tr[ \rho |\Psi\rangle\langle \Psi|]>\lambda_1$. Take $\sigma=|\Psi\rangle\langle \Psi|$. We claim that
	\begin{equation}\label{eq:thm4less}
		\min\{\tr[\rho_A\sigma_A],\tr[\rho_B\sigma_B]\}\leq \lambda_1.
	\end{equation}
	In fact, as
	$ \sigma_A= \sum_{i=1}^l \lambda_i |e_i\rangle\langle e_i|$ and $\sigma_B= \sum_{i=1}^l \lambda_i |f_i\rangle\langle f_i|$, we have
\begin{equation}
\tr[\rho_A \sigma_A]=\sum_{i=1}^l \lambda_i \langle e_i| \rho_A |e_i\rangle \leq  \sum_{i=1}^l \lambda_1 \langle e_i |\rho_A |e_i\rangle \leq  \lambda_1 \tr[\rho_A]\leq \lambda_1
\end{equation}
	and
	\begin{equation}\tr[\rho_B \sigma_B]=\sum_{i=1}^l \lambda_i\langle f_i| \rho_B |f_i\rangle \leq  \sum_{i=1}^l \lambda_1 \langle f_i |\rho_B |f_i\rangle \leq  \lambda_1 \tr[\rho_B]\leq \lambda_1.\end{equation}
	Hence,  by the inequality \eqref{eq:thm4less}, we have the following inequalities,
	\begin{equation} \langle \rho, \sigma \rangle=\tr[\rho \sigma ] >\lambda_1\geq  \min\{\langle \rho_A,\sigma_A\rangle,\langle \rho_B,\sigma_B\rangle\},\end{equation}
	which implies that $\rho \in   \mathcal{E}_{\mathrm{IPC}}.$ Therefore, we have
	$\mathcal{E}_{\mathrm{FBC}}  \subseteq   \mathcal{E}_{\mathrm{IPC}}$.
\end{proof}

   	Now we give some analysis of the entanglement detectability of $\rho(x)$, 
   	\begin{equation}
   		\rho(x)=(1-x)\frac{ \mathbb{I}_{(d-1)^2}}{(d-1)^2} + x|\Psi\rangle\langle \Psi|
   	\end{equation}
   	in Example 2 via IPC,   PC, FBC, and $p_3$-PPT criteria.   	These states are   separable if and only if   $0<x\leq 1.$   We show that the entanglement $\rho(x)$ can be also detected via reduction criterion (hence IPC) below.   If $x>\frac{d-2}{d^2-d-1},$ then  one may apply fidelity based criterion  corresponding to $\mathcal{W}_{\mathrm{FBC}}(\Psi) $ 
   	to detect the   entanglement of $\rho(x)$ (the states corresponding to $(d, x)$ values are plotted  above the  dashed  blue line in Fig.  \ref{fig:Example2}).  
   	However, if  the maximal eigenvalue  of $\rho(x)$, denote as $\Delta=\frac{m+nd+\sqrt{(m+nd)^2-4mn}}{2} $ ($m=\frac{1-x}{(d-1)^2}$ and $n=\frac{x}{d}$),  is less than $\frac{1}{d}$, then the fidelity based criterion fails to detect the entanglement (the states corresponding to $(d, x)$ values are plotted  by the blue area in Fig.    \ref{fig:Example2}). Concerning the entanglement detectability of PC via inequality \eqref{eq:rhox_PC}, we have strict inclusion,
   	\begin{equation}
   		\left(\mathcal{E}_{\mathrm{PC}}\cup \mathcal{E}_{\mathrm{FBC}}\right) \subsetneqq\mathcal{E}_{\mathrm{IPC}} .
   	\end{equation}
   	Therefore, our entanglement criterion IPC performances better than the well known experimentally accessible purity and fidelity-based criteria.
   	
   	Now we compare the entanglement detectability between IPC and $p_3$-PPT criteria. One finds that

   	\begin{equation}\label{eq:PT_p3}
   		\begin{array}{rcl}
   			p_2^2-p_3&=&\displaystyle \frac{x}{(d-1)^4 d^2}  \Big [\left(d^3-2 d^2+2\right)^2 x^3\\[4mm]
   			& &+\left(2 d^4-12 d^3+18 d^2-5 d-6\right) x^2\\[2mm]
   			& &+\left(-d^4+8 d^3-15 d^2+10 d+1\right) x-d \Big].
   		\end{array} 
   	\end{equation} 
   	For small enough $x$,  the right handside of Eq. \eqref{eq:PT_p3} is negative, i.e.,     $p_2^2<p_3$. So  $p_3$-PPT  criterion fail to detect the entanglement of $\rho(x)$ for such $x$ (the states corresponding to $(d, x)$ values are plotted below the dotted blue line in Fig. \ref{fig:Example2}).   The detail analysis can be found below.  
   	
   	\begin{enumerate}
   	\item [\rm(A)] The entanglement of the  state $\rho(x)$ in Example 2 can be  detected via IPC or RC  if and only if  $0<x\leq 1.$
   	
   	It is sufficient to show that
   	$ \rho(x)_A\otimes \mathbb{I}_B-\rho(x)$
   	is not positive semidefinite.
   	In fact, $ \rho(x)_A=\left(\frac{1-x}{(d-1)} \mathbb{I}_{d-1} +\frac{x}{d} \mathbb{I}_A\right)$.
   	Therefore, $\rho(x)_A\otimes \mathbb{I}_B-\rho(x)$ equals to
   	\begin{equation}
   	\left(\frac{1-x}{(d-1)} \mathbb{I}_{d-1} +\frac{x}{d} \mathbb{I}_A\right)\otimes \mathbb{I}_B- (1-x)\frac{ \mathbb{I}_{(d-1)^2}}{(d-1)^2} - x|\Psi\rangle\langle \Psi|.\end{equation}
   	Notice that the minor with rows and columns corresponding to $\{|11\rangle,|dd\rangle \}$ is as follows,
   	\begin{equation}
   	\left[\begin{array}{cc}
   		\displaystyle\frac{(d-2)(1-x)}{ (d-1)^2 }& -\displaystyle\frac{x}{d}\\[3mm]
   		-\displaystyle\frac{x}{d} & 0
   	\end{array}\right],
   	\end{equation}
   	whose determinant is $-\frac{x^2}{d^2}<0$ wherever $x>0$. So we can conclude that $ \rho(x)_A\otimes \mathbb{I}_B-\rho(x)$ is not positive semidefinite wherever $0<x\leq 1.$

   			\item [\rm(B)] The entanglement of the  state $\rho(x)$ in Example 2 can be  detected via PC  if and only if
   		\begin{equation}\label{eq:rhox_PC}
   			\frac{(1-x)^2}{(d-1)^2} +x^2+  \frac{2(1-x)x}{(d-1)d}>\frac{(1-x)^2}{(d-1) } +\frac{x^2}{d}+\frac{2(1-x)x}{d}.
   		\end{equation}
   		
   		On the one hand, $\tr[(\rho(x))^2]$ is given by
   		\begin{equation}\frac{(1-x)^2}{(d-1)^2} +x^2+2\times \frac{(1-x)x}{(d-1)^2} \frac{d-1}{d}=\frac{(1-x)^2}{(d-1)^2} +x^2+  \frac{2(1-x)x}{(d-1)d}.\end{equation}
   		On the other hand, the $\tr[(\rho(x)_A)^2]$ is equal to
   		\begin{equation}\frac{(1-x)^2}{(d-1)} +\frac{x^2}{d}+2\times \frac{(1-x)x}{(d-1)d} (d-1)  =\frac{(1-x)^2}{(d-1) } +\frac{x^2}{d}+\frac{2(1-x)x}{d} .\end{equation}
   		By PC and the above two equations, we must have the inequality \eqref{eq:rhox_PC} hold if the entanglement of $\rho(x)$ can be detected by PC  (the states corresponding to $(d, x)$ values are plotted  above the dotdashed blue line in Fig. \ref{fig:Example2}).

   	 	\item [\rm(C)] The entanglement of the  state $\rho(x)$ in Example 2 can be  detected  by  FBC  via witness $\mathcal{W}_{\mathrm{FBC}}(\Psi) $ if $x>\frac{d-2}{d^2-d-1}$. This can be easily deduced once one notices that
   	 $\langle  \Psi| \rho(x)|\Psi\rangle =\frac{1-x}{d(d-1)}+x.$
   	 
   	 Moreover, by  Proposition 1  and the maximal eigenvalue of $\rho(x)$ given in Appendix C, we have that if $\Delta\leq \frac{1}{d}$, then the entanglement of $\rho(x)$ can not be detected by any witnesses arising from FBC.

   	 \item [\rm(D)] The entanglement of the  state $\rho(x)$ in Example 2 can be detected via $p_3$-PPT criterion if and only if
   	 \begin{equation} \left(d^3-2 d^2+2\right)^2 x^3+\left(2 d^4-12 d^3+18 d^2-5 d-6\right) x^2+\left(-d^4+8 d^3-15 d^2+10 d+1\right) x-d>0.\end{equation}
   	 
   	 To compare the IPC with the PT moments approach: $ p_2^2\leq  p_3$ for separable $\rho$, where $p_k=\tr[ (\rho^{T_A})^k]$, we calculate the value $p_2$ and $p_3$ for $\rho(x)$. We have
   	 \begin{equation} \rho(x)^{T_A}=(1-x)\frac{ \mathbb{I}_{(d-1)^2}}{(d-1)^2} + x (|\Psi\rangle\langle \Psi|)^{T_A}\end{equation}
   	 and
   	 \begin{equation}(|\Psi\rangle\langle \Psi|)^{T_A}=\frac{1}{d}\sum_{i\leq j} |v_{ij}\rangle\langle v_{ij}|-\frac{1}{d} \sum_{k< l} |w_{kl}\rangle\langle w_{kl}|,
   	 \end{equation}
   	 where $|v_{ij}\rangle=(|ij\rangle+|ji\rangle)/2$ and $|w_{kl}\rangle=(|kl\rangle-|lk\rangle)/2.$
   	 By direct calculation we have
   	 \begin{equation}p_2^2-p_3=\frac{x}{(d-1)^4 d^2} \left[\left(d^3-2 d^2+2\right)^2 x^3+\left(2 d^4-12 d^3+18 d^2-5 d-6\right) x^2+\left(-d^4+8 d^3-15 d^2+10 d+1\right) x-d\right].\end{equation} 
   	The   corresponding line $p_2^2=p_3$ is plotted   as the dotted blue line in Fig. \ref{fig:Example2}.	For small enough $x$,  the number  $p_2^2-p_3<0.$  
   	 	
   	\end{enumerate}

\section{Appendix E:  IPC for multipartite entanglement}

\emph{Inner product criterion for multipartite entanglement.---} Our inner product criterion can be also generalized to detect multipartite entanglement.

\emph{Proposition 2.---} (IPC for multipartite entanglement) Let $\rho,\sigma\in \mathbb{D}(\otimes_{i=1}^n\mathcal{H}_{A_i})$ be      multipartite density matrices. Denote $\mathcal{A}=\{A_1,A_2,\cdots,A_n\}$. Let $\mathcal{P}(\mathcal{A})$ be the set of all possible bipartitions $S|\overline{S}$ of $\mathcal{A}$, where $\overline{S}:=\mathcal{A}\setminus\{S\}.$
If
\begin{equation}\label{eq:Inner_ineq_mul}
	\langle \rho,\sigma\rangle > \min_{(S|\overline{S})\in \mathcal{P}(\mathcal{A})} \ \min\{\langle \rho_S,\sigma_S\rangle, \langle \rho_{\overline{S}},\sigma_{\overline{S}}\rangle \} ,
\end{equation}
then both $\rho$ and $\sigma $ are entangled.

\begin{proof} First, if $\rho$ is entangled, i.e., fully separable, then $\rho$ can be written as
	\begin{equation}\label{eq:mul_ent_full} \rho =\sum_{i} p_i \rho_{i}^{A_1} \otimes  \rho_{i}^{A_2} \otimes  \cdots \otimes   \rho_{i}^{A_n}.
	\end{equation}
	Then for each bipartition $S|\overline{S}\in \mathcal{P}(\mathcal{A})$,   $\rho $ can be written as
	\begin{equation}\label{eq:mul_ent_bipartite}\rho =\sum_{i} p_i \rho_{i}^{S} \otimes  \rho_{i}^{\overline{S}},
	\end{equation}
	where $\rho_{i}^{S}=\otimes_{A_j\in S}\rho_{i}^{A_j}$ and $\rho_{i}^{\overline {S}}=\otimes_{A_j\in \overline{S}}\rho_{i}^{A_j}.$ Hence by Theorem 1 we have
	\begin{equation}\label{eq:mul_ent_bi_ineq}
		\langle \rho , \sigma\rangle \leq  \min\{\langle \rho _{S}, \sigma_S \rangle, \langle \rho _{\overline{S}}, \sigma_{\overline{S}}  \rangle  \}.
	\end{equation}
	Therefore,
	\begin{equation}\label{eq:mul_ent_bi__mul_ineq}
		\langle \rho , \sigma\rangle \leq \min_{S|\overline{S}\in \mathcal{P}(\mathcal{A})}  \min\{\langle \rho _{S}, \sigma_S \rangle, \langle \rho _{\overline{S}}, \sigma_{\overline{S}}  \rangle  \},
	\end{equation}
	which is contradicted with the given condition. Therefore, the state $\rho$ must be entangled. So is the state $\sigma$.
\end{proof}

\emph{Example 4.---} We consider the $n$-qubit noisy GHZ states,
\begin{equation}
\rho_n(p)=p |\mathrm{GHZ}_n\rangle \langle \mathrm{GHZ}_n|+ (1-p) \frac{\mathbb{I}_{2^n}}{2^n}, \end{equation}
where $|\mathrm{GHZ}_n\rangle=\frac{1}{\sqrt{2}}(|0\rangle^{\otimes n}+|1\rangle ^{\otimes n}).$
If we set $\sigma$ to be the state $|\mathrm{GHZ}_n\rangle\langle \mathrm{GHZ}_n|,$ we get $\langle \rho_n(p), \sigma\rangle = \frac{1-p}{2^{n}}+ p$. For each bipartition $S|\overline{S}$ of $\{A_1,A_2,\cdots, A_n\}$, we have $  \langle \rho_S,\sigma_S\rangle =\frac{1-p}{2^{k}}+ \frac{p}{2}$ and $  \langle \rho_{\overline{S}},\sigma_{\overline{S}}\rangle =\frac{1-p}{2^{n-k}}+ \frac{p}{2}$, where $|S|=k$. Hence the lower bound in \eqref{eq:Inner_ineq_mul} is
$\frac{1-p}{2^{n-1}}+ \frac{p}{2}$, and $\rho_n(p )$ is entangled if $p>\frac{1}{2^{n-1}+1}$.

\vskip 5pt
 
It has been pointed out that  the universal state inversion map \cite{Eltschka18b}
\begin{equation} \mathcal{I}=\left[ \prod_{j=1}^n \left(\mathrm{Tr}_j(\cdot ) \otimes\mathds{1}_j- \mathrm{id} \right)\right]\end{equation}
or the even much more  generalize $T$-inversion map (where $T\subseteq \{1,2,3,\cdots, n\}$ and $0\leq \alpha_j,\beta_k\leq 1$) \cite{Eltschka18}  
\begin{equation} \mathcal{I}_T^{\{\alpha_j,\beta_k\}} =\prod_{j\in T}\left[   \mathrm{Tr}_j(\cdot ) \otimes\mathds{1}_j- \alpha_j\mathrm{id} \right] \prod_{k\notin T}\left[   \mathrm{Tr}_k(\cdot ) \otimes\mathds{1}_j+\beta_k\mathrm{id} \right]\end{equation}
is a  positive but not completely positive map. Hence, these maps could be used to detect entanglement. In fact, the $r$-reduction  criterion can be seen as the derivation from the positive but not completely positive   of the map \begin{equation}\Lambda_{-\frac{1}{r}}=  \mathrm{Tr}_A(\cdot ) \otimes\mathds{1}_A-\frac{1}{r}  \mathrm{id}_A \end{equation}
and  the nonpositive of $\Lambda_{-\frac{1}{r}}\otimes \mathrm{id}_B (\rho)= \mathds{1}_A \otimes \rho_B-\frac{1}{r} \rho$
implies that the Schmidt number of $\rho$ must no less than $r$. In the following, we would like to extend similar argument to multipartite system.

First, we consider the   map
\begin{equation}\Lambda^A_{-\frac{1}{r}} \otimes \Lambda^B_{1}=\left[\mathrm{Tr}_A(\cdot ) \otimes\mathds{1}_A-\frac{1}{r}  \mathrm{id}_A \right] \otimes \left[\mathrm{Tr}_B(\cdot ) \otimes\mathds{1}_B+ \mathrm{id}_B \right] .\end{equation}
By Ref. \cite{Eltschka18}, this is a positive map. To use it to detect entanglement, we should introduce another system call Charlie $C$ and consider the map $\Lambda=\Lambda^A_{-\frac{1}{r}} \otimes \Lambda^B_{1} \otimes \mathrm{id}_C.$
For any $\rho_{AC}$ with $\mathrm{SN}(\rho_{AC})\leq r$, by the $r$-reduction criterion, $\Lambda^A_{-\frac{1}{r}} \otimes \mathrm{id}_C (\rho_{AC}) \geq \mathbf{0}.$ And by the positivity  of $\Lambda^B_{1}$, we have for any state $\rho_B$ of subsystem $B$, 
\begin{equation}\label{eq:AC_B}
	\Lambda(\rho_{AC} \otimes \rho_B) = \left(\Lambda^A_{-\frac{1}{r}} \otimes \mathrm{id}_C\right) (\rho_{AC})  \otimes \Lambda_1^B(\rho_B) \geq \mathbf{0}.
\end{equation} 
Now for any state $\rho_{AB}$ of subsystems $AB$ and $\rho_C$ of subsystem $C$, then by the positivity of $\Lambda^A_{-\frac{1}{r}} \otimes \Lambda^B_{1}$ and $\mathrm{id}_C$, we have 
 \begin{equation}\label{eq:AB_C}
 	\Lambda(\rho_{AB} \otimes\rho_C) = \left(\Lambda^A_{-\frac{1}{r}} \otimes \Lambda^B_{1}\right) (\rho_{AB})  \otimes \rho_C \geq \mathbf{0}.
 \end{equation}
And for any state $\rho_{BC}$ of subsystems $BC$ and $\rho_A$ of subsystem $A$, then by the positivity of $ \Lambda^B_{1}\otimes \mathrm{id}_C$ and $\Lambda^A_{-\frac{1}{r}}$, we have 
 \begin{equation}\label{eq:BC_A}
	\Lambda(\rho_{BC} \otimes \rho_A) = \left(\Lambda^B_{1}\otimes \mathrm{id}_C\right) (\rho_{BC})  \otimes  \Lambda^A_{-\frac{1}{r}}( \rho_A )\geq \mathbf{0}.
\end{equation}

From the above  Eqs.  \eqref{eq:AC_B}, \eqref{eq:AB_C} and  \eqref{eq:BC_A}, we can arrive at the following conclusion.

\vskip 5pt

\emph{Proposition 3.---} Let $\Lambda=\Lambda^A_{-\frac{1}{r}} \otimes \Lambda^B_{1} \otimes \mathrm{id}_C.$ If $\rho,\sigma\in  \mathbb{D}(\mathcal{H}_A\otimes \mathcal{H}_B\otimes \mathcal{H}_C )$  and $\langle \Lambda(\rho), \sigma\rangle=\mathrm{Tr}[\Lambda(\rho) \sigma] <0$, then
$\rho$ is either has genuine entanglement or for each form  bipartite decomposition of $\rho$ 
\begin{equation}\rho=\sum_{S\in \{A,B,C\} } p_{S|\overline{S}} \left(\sum_{i} p_{S,i} \rho_{S,i} \otimes \rho_{\overline{S},i}\right),\end{equation}
 there always exists some  $\rho_{AC,i}$ whose Schmidt number $\mathrm{SN}(\rho_{AC,i})>r.$ Moreover, the statement also holds for $\sigma.$

The last statement holds because
 \begin{equation}
	\langle \Lambda(\rho), \sigma\rangle= \langle \rho_C, \sigma_C\rangle +\langle \rho_{BC}, \sigma_{BC}\rangle- \frac{1}{r}\langle \rho_{AC}, \sigma_{AC}\rangle-\frac{1}{r}\langle \rho , \sigma \rangle=\langle \rho,\Lambda( \sigma)\rangle.
\end{equation}

\vskip 5pt

\emph{Example 5.---} We consider the $3$-qudit noisy GHZ states,
\begin{equation}
\rho_n(p)=p |\mathrm{GHZ} \rangle \langle \mathrm{GHZ} |+ (1-p) \frac{\mathbb{I}_{d^3}}{d^3}, \end{equation}
where $|\mathrm{GHZ}\rangle=\frac{1}{\sqrt{d}}\sum_{j=0}^{d-1} |jjj\rangle.$
If we set $\sigma$ to be the state $|\mathrm{GHZ}\rangle\langle \mathrm{GHZ}|,$ we get 
\begin{equation}\langle \Lambda\left( \rho_n(p)\right), \sigma\rangle = p\left(\frac{2}{d}-\frac{1}{r}(\frac{1}{d}+1)\right)+(1-p).\end{equation}
One finds that if $r<\frac{d+1}{\frac{1-p}{p}d+2}$, then $\langle \Lambda\left( \rho_n(p)\right), \sigma\rangle<0.$ Let $r_{\mathrm{op}}$ be the maximal integer such that $r_{\mathrm{op}}<\frac{d+1}{\frac{1-p}{p}d+2}$ holds.  Then $\rho_n(p)$ is either genuine entangled or each its bipartite decomposition contains three terms $\rho_{AB,i},\rho_{BC,j}$ and $\rho_{AC,k}$ such that all of them are of Schmidt number greater than $r_{\mathrm{op}}.$

\end{document}